\documentclass[usenatbib]{mnras}
\usepackage{epsfig}
\usepackage{times}
\usepackage{graphicx}
\usepackage{amsmath}

\def\Hy@FixNotFirstPage{%
	\gdef\Hy@FixNotFirstPage{%
		\setbox\AtBeginShipoutBox=\hbox{%
			\copy\AtBeginShipoutBox
		}%
	}%
}
\AtBeginShipout{\Hy@FixNotFirstPage}
 
\def\I{\,\textsc{i}}
\def\II{\,\textsc{ii}}
\def\III{\,\textsc{iii}}
\def\kms{\text{km s}^{-1}}

\def\Mdot{\,\mathrm{M}_{\odot}}
\def\days{\,\mathrm{days}}
\def\mags{\,\mathrm{mag}}

\title[Study of the Type Ia/IIn Supernova PS15si]{An Optical and Near-Infrared Study of the Type Ia/IIn Supernova PS15si}

\author[Kilpatrick et al.]{Charles D. Kilpatrick$^1$\thanks{Email:
    charlesk@email.arizona.edu}, Jennifer E. Andrews$^1$, Nathan Smith$^1$, Peter Milne$^1$,
	\newauthor George H. Rieke$^1$, WeiKang Zheng$^2$, Alexei V. Filippenko$^2$ \\$^1$Steward
  Observatory, University of Arizona, Tucson, AZ 85721, USA \\$^2$Department of Astronomy, University of California, Berkeley, CA 94720-3411, USA}

\begin{document}
\date{Accepted 0000, Received 0000, in original form 0000}
\pagerange{\pageref{firstpage}--\pageref{lastpage}} \pubyear{2015}
\maketitle
\label{firstpage}

\begin{abstract}
\noindent
We present optical/near-infrared spectroscopy and photometry of the supernova (SN) PS15si. This object was originally identified as a Type IIn SN, but here we argue that it should be reclassified as a Type Ia SN with narrow hydrogen lines originating from interaction with circumstellar matter (CSM; i.e., SN~Ia/IIn or SN~Ia-CSM). Based on deep nondetections $27\days$ before discovery, we infer that this SN was discovered around or slightly before optical maximum, and we estimate the approximate time that it reached $R$-band maximum based on comparison with other SNe~Ia/IIn. In terms of spectral morphology, we find that PS15si can be matched to a range of SN~Ia spectral types, although SN~1991T-like SNe~Ia provides the most self-consistent match. While this observation agrees with analysis of most other SNe~Ia/IIn, we find that the implied CSM luminosity is too low to account for the overall luminosity of the SN at a time when the CSM should outshine the underlying SN by a few magnitudes.  We infer that the similarity between PS15si and the hot, overluminous, high-ionisation spectrum of SN~1991T is a consequence of a spectrum that originates in ejecta layers that are heated by ultraviolet/X-ray radiation from CSM interaction. In addition, PS15si may have rebrightened over a short timescale in the $B$ and $V$ bands around $85\days$ after discovery, perhaps indicating that the SN ejecta are interacting with a local enhancement in CSM produced by clumps or a shell at large radii.

\end{abstract}


%%%%%%%%%%%%%%%%%%%%%%%%%%%%%%%%%%%%%%%%%%%%%%%%%%%%%%%%%%%%%%%%%%%%%%
\section{INTRODUCTION}\label{sec:intro}

The increasing number of targets from high-cadence surveys has revealed enormous variety in spectroscopic and photometric signatures of supernovae (SNe) including the subclass of Type Ia SNe (SNe~Ia). While some variations in SN~Ia spectra and light curves have long been recognised, such as the width-luminosity or Phillips relation \citep{phillips93}, recent focus on ejecta velocities \citep{wang+09}, spectral variation with time \citep{patat+07}, and polarization signals \citep{kasen+03,patat+09,porter+16} suggest a wide variety of SN~Ia subtypes. Peculiar trends in these measurements contribute to the ambiguity in the underlying mechanism(s) for SNe~Ia. Despite the success of the C/O white dwarf (WD) thermonuclear explosion model in explaining the explosion mechanism of SNe~Ia \citep{hf60,arnett68,nomoto86}, it is still unclear whether a dominant evolutionary channel exists to ignite these explosions. Two models predict different channels through which ignition might occur --- a single-degenerate model, involving a WD that accretes from a nondegenerate companion star, and a double-degenerate model, involving the merger of two WDs. Various lines of evidence from spectroscopic and photometric signatures of various SNe~Ia sub-types seem to argue in favor of both models. For example, it has been suggested that helium in peculiar SNe~Iax \citep{foley+13} originates from nondegenerate companion stars, and signatures of circumstellar matter (CSM) in thermal X-ray emission and spectropolarimetry in some SNe~Ia also point toward mass loss from a nondegenerate companion \citep{wang+96,immler+06,patat+12}. At the same time, so-called super-Chandrasekhar SNe exhibit high luminosity and low ejecta velocities, and point toward massive ($\sim 2~\Mdot$) WD progenitors, possibly from WD mergers \citep{howell+06,hicken+07,silverman+11}. While illuminating the diversity of SNe~Ia, these systems generate additional questions. Do they represent the extremes along a continuum of SN~Ia explosion scenarios?  What similarities do these SNe~Ia share with each other and the more common classifications?

In particular, one new class of SN~Ia-like events exhibits the spectroscopic signatures of both SNe~Ia and SNe~IIn (IIn for narrow lines of hydrogen) in the form of broad Fe, Ca, S, and Si absorption combined with strong, narrow H$\alpha$ emission, consistent with a SN~Ia explosion encountering dense CSM. The dominant hydrogen feature seen in this subclass contrasts with the majority of SNe~Ia where H$\alpha$ searches have yielded null results to deep limits of $\sim0.001-0.01~\Mdot$ \citep[e.g., SNe 2005am, 2005cf, and 2011fe as in][]{leonard07,shappee+13}. The first SN identified with both SN~Ia and SN~IIn features, SN~2002ic \citep{hamuy+03}, exhibited absorption features characteristic of SN~1991T-like SNe~Ia \citep{filippenko+92a,phillips+92,filippenko97}, but was identified as having both broad and narrow H$\alpha$ profiles as early as $6\days$ after maximum light. SN~2002ic and SNe with similar spectroscopic signatures have been denoted as SNe~Ia/IIn, IIa, and sometimes Ia-CSM\footnote{Hereafter, we refer to this class of objects as SNe~Ia/IIn, since the name SN~Ia-CSM mixes an interpretation of a physical mechanism with a spectral classification.} \citep{deng+04,kotak+04,silverman+13} owing to their spectroscopic overlap with both SNe~Ia and IIn. Subsequent to the discovery of SN~2002ic, two SNe~IIn --- SNe 1997cy and 1999E \citep{hamuy+03,wv04} --- were reclassified as SN~2002ic-like. More recently, several SNe~Ia/IIn, such as SN~2005gj and PTF11kx \citep{aldering+06,prieto+07,dilday+12}, were spectroscopically identified soon after explosion. It is now estimated that this class represents as many as $0.1-1\%$ of all SNe~Ia \citep{dilday+12}.

In addition to their unique spectroscopic properties, the light curves of SNe~Ia/IIn also exhibit traits that overlap with both SNe~Ia and SNe~IIn. SNe~Ia/IIn consistently have $R$-band peak luminosities brighter than $-19\mags$, and their light curves evolve slowly with a linear decay that can last for several weeks. For the first $25\days$ after peak, SNe~Ia/IIn light curves are generally consistent with ``normal'' SNe~Ia \citep{hamuy+03,prieto+07}. After this point, SNe~Ia/IIn decline much more slowly than SNe~Ia, and the residual flux is assumed to be due to CSM interaction. In this model, the freely expanding, $^{56}$Ni-powered SN can be observed mostly at early times and the optically thick CSM interaction is dominant at later times \citep{cy04}, perhaps followed by a return to the $^{56}$Ni-powered decay line.

While the general trend of slow decay is observed, it is still an open question why some SNe~Ia/IIn exhibit strong SN~Ia-like absorption in early-time spectra \citep[e.g.,][]{hamuy+03,silverman+13} while others appear to be diluted by a thermal continuum \citep[e.g.,][]{aldering+06}. During the period when H$\alpha$ emission fades (generally $\sim$70--300 days), the spectrum further evolves as the thermal continuum becomes less apparent and the SN~Ia-like component enters the nebular phase. This epoch in SN~Ia/IIn evolution is critical for determining the characteristics of both the CSM and the intrinsic SN emission where competing models of the progenitor system can be evaluated \citep{chugai+04}. This phase also probes the CSM out to large radii where observations may reveal complex structure in the surrounding medium.

In this paper we discuss PS15si, discovered by the Pan-STARRS Survey for Transients \citep{atel7280} on 2015 Mar. 23 (all dates presented herein are UT) and located 1\farcs6 from the center of the galaxy 2dFGRS N166Z116 (hereafter N166Z116). The target was originally classified $5\days$ after discovery as a SN~IIn, with narrow hydrogen lines indicating redshift $z=0.053$ \citep{atel7308}. In this paper, we argue that PS15si is a SN~Ia/IIn. We adopt a Milky Way line-of-sight reddening of $E(B-V)=0.046\mags$ \citep{sf11}, a distance to N166Z116 of $219.4 \pm 15.4~\text{Mpc}$, and $m-M=36.71 \pm 0.15\mags$ \citep{colless+03}. From the light curve and prediscovery images\footnote{The target was provided from the Pan-STARRS NEO survey and made public via \url{star.pst.qub.ac.uk/ps1threepi/}.}, we infer that the SN was discovered around or slightly before optical maximum given the discovery magnitude $w=17.16$ (Sloan $g+r+i$) on 2015 Mar. 23 \citep[see][]{huber+15}.  Assuming a line-of-sight $A_{w}=0.125$ mag, the peak apparent magnitude indicates PS15si had an absolute $w$-band magnitude of $M_{w} \approx -19.7\mags$.

\section{OBSERVATIONS}\label{sec:obs}

\begin{figure}
	\includegraphics[width=0.475\textwidth]{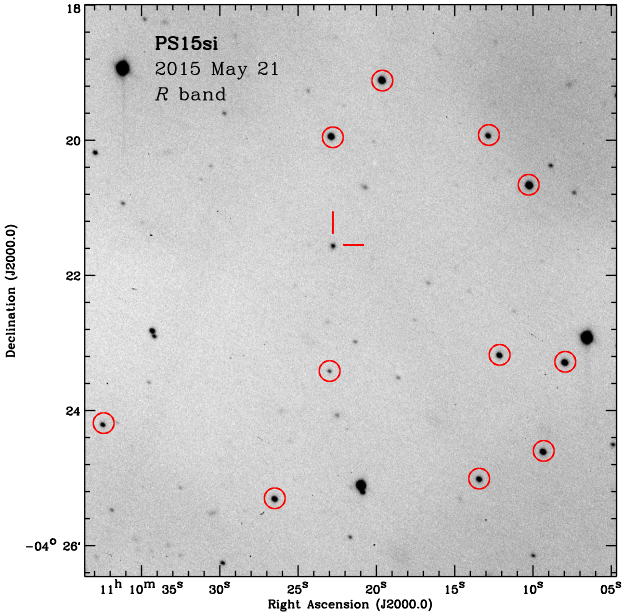}
	\caption{$R$-band image of PS15si obtained on 2015 May 21 by Super-LOTIS. The position of the SN is indicated along with stars used to perform differential photometry (as described in \autoref{sec:obs}). The SN was at $R = 17.51\mags$ at this time.}\label{fig:example}
\end{figure}

We took optical imaging photometry of PS15si using the Super-LOTIS \citep[Livermore Optical Transient Imaging System;][]{williams+08} 0.6~m telescope at Kitt Peak National Observatory between 2015 May 8 and Jun. 20. Standard reductions, including flat-fielding and bias subtraction, were carried out using a semi-automatic routine. We then performed differential photometry using stars in the field and APASS standards. A Super-LOTIS image including the SN from 2015 May 21 is shown in \autoref{fig:example}. We transformed $R$ and $I$-band standards from APASS $r'$ and $i'$ magnitudes as described by \citet{jester+05}. 

\begin{figure*}
\includegraphics[width=0.49\textwidth]{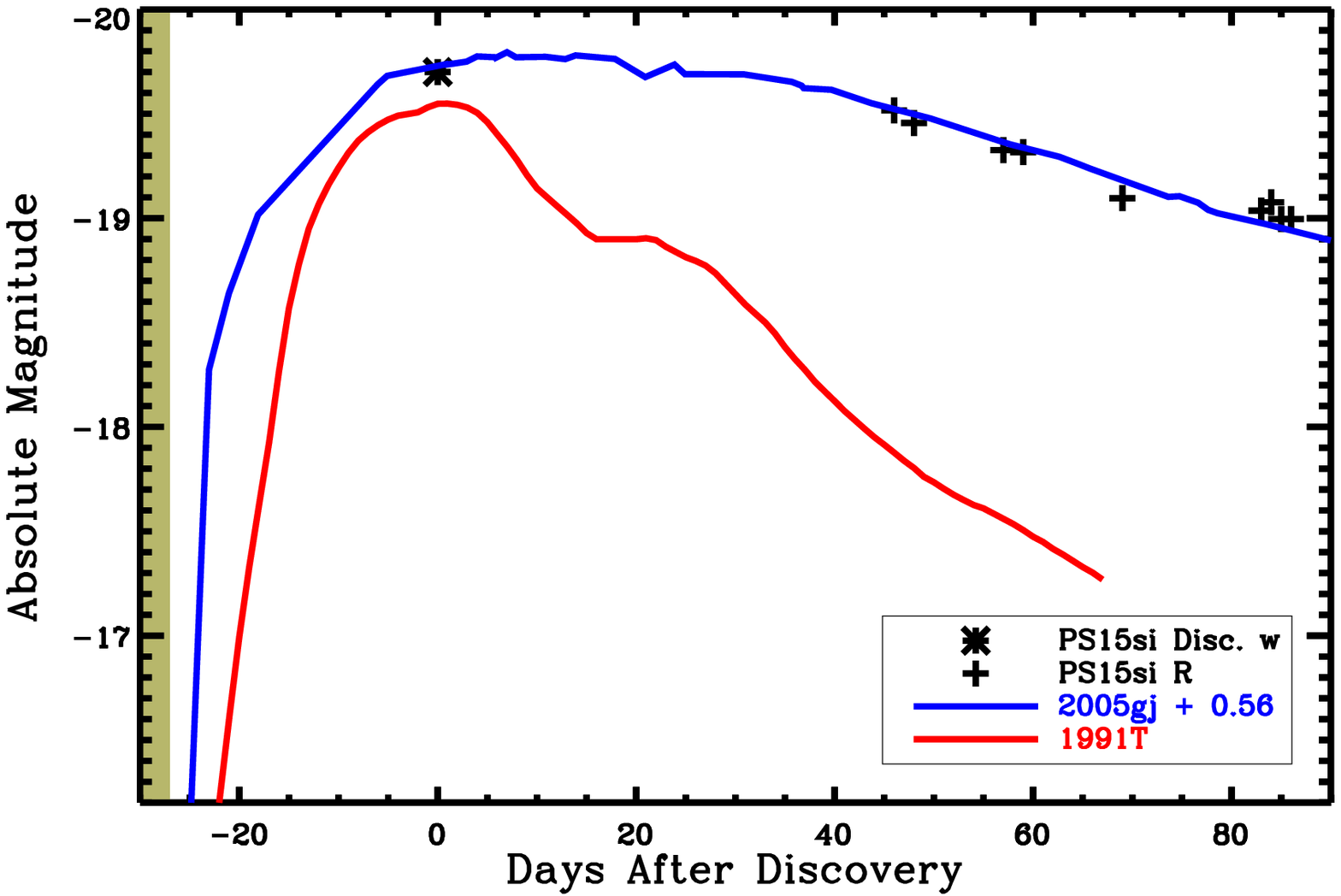}
\includegraphics[width=0.49\textwidth]{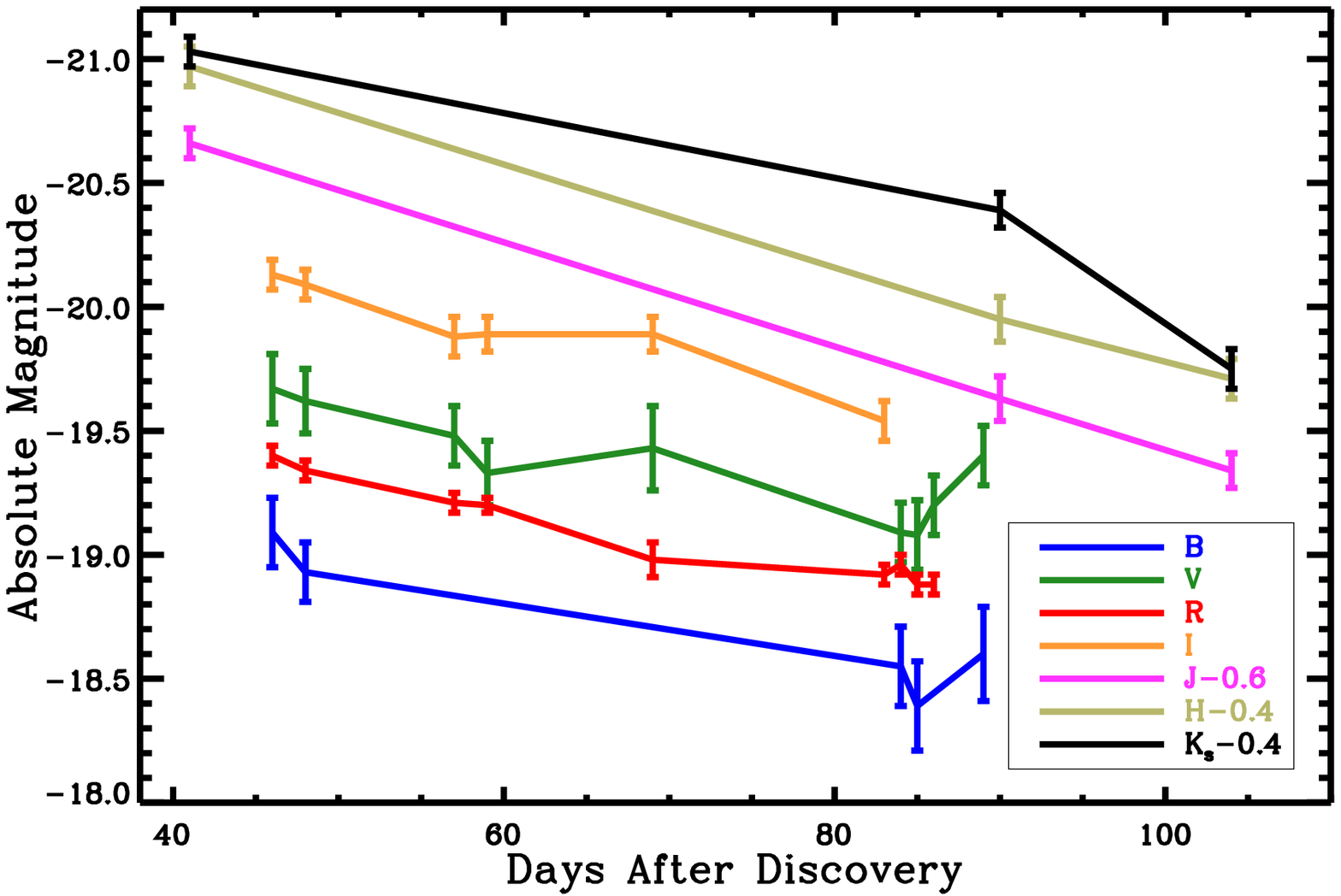}
\caption{(Left) $R$-band photometry of PS15si is plotted as crosses along with the discovery magnitude (in the $w$ band; i.e., $g+r+i$). The shaded region at the left around $-27\days$ corresponds to times when nondetections were reported in deep, prediscovery photometry (as described in \autoref{sec:date}). We show the $R$-band light curve of a SN~1991T-like (SN~Ia) template in red derived from models in \citet{stern+04}. We also plot $R$-band photometry of the known SN~Ia/IIn 2005gj in blue, which has been dimmed by $0.56\mags$ (without any stretch) for comparison to the $R$-band decline of PS15si. The phase of the SN~2005gj and SN~1991T-like light curves (i.e., in time since explosion) are matched to each other. (Right) Our BVRIJHK$_{s}$ photometry of PS15si.}\label{fig:phot}
\end{figure*}

We took near-infrared (NIR) observations with the 3.8~m United Kingdom Infrared Telescope (UKIRT) on Mauna Kea using WFCAM2 (2015 May 3, Jun. 21, Jul. 5)\footnote{Our observing campaign for PS15si ended at this point due to solar conjunction.}. $JHK$ observations were pipeline reduced by the Cambridge Astronomical Survey Unit (CASU). We performed aperture photometry on all images using a $1\farcs3$ diameter aperture to reduce contamination from the host galaxy, and we calibrated the instrumental magnitudes using 2MASS $JHK_{s}$ NIR standard stars present in the field. For both optical and NIR magnitudes, we calculated uncertainties by adding in quadrature photon statistics and zero-point deviation of the standard stars for each epoch. We detail the optical and NIR magnitudes for PS15si in \autoref{tab:phot} and \autoref{fig:phot}.

\begin{table}
\begin{center}\begin{minipage}{3.3in}
      \caption{Optical and UKIRT Photometry of PS15si}\scriptsize
\begin{tabular}{@{}cccccc}\hline\hline
  UT Date & day\footnote{Since discovery on 2015 Mar. 23.}& $B$ & $V$ & $R$ & $I$ \\ %\hline       
  (y-m-d) & & & & &    \\   \hline
  2015-05-08&46	& 17.62$\pm$0.14	& 17.04$\pm$0.14	& 17.31$\pm$0.04	& 16.58$\pm$0.06 	\\
  2015-05-10&48	& 17.78$\pm$0.12	& 17.09$\pm$0.13	& 17.37$\pm$0.04	& 16.62$\pm$0.06 	\\
  2015-05-19&57	& ...				& 17.23$\pm$0.12	& 17.50$\pm$0.04	& 16.83$\pm$0.08 	\\
  2015-05-21&59	& ...				& 17.38$\pm$0.13	& 17.51$\pm$0.03	& 16.82$\pm$0.07 	\\
  2015-05-31&69	& ...				& 17.28$\pm$0.17	& 17.73$\pm$0.07	& 16.82$\pm$0.07 	\\
  2015-06-14&83	& ...				& ...				& 17.79$\pm$0.04	& 17.17$\pm$0.08 	\\
  2015-06-15&84	& 18.16$\pm$0.16	& 17.62$\pm$0.12	& 17.75$\pm$0.04	& ...				\\
  2015-06-16&85	& 18.32$\pm$0.18	& 17.63$\pm$0.14	& 17.83$\pm$0.04	& ...				\\
  2015-06-17&86	& ...				& 17.51$\pm$0.12 	& 17.83$\pm$0.04	& ...				\\
  2015-06-20&89	& 18.11$\pm$0.19	& 17.31$\pm$0.12	& ...				& ...				\\
  \hline
  UT Date &day& $J$ & $H$ & $K_{s}$ \\ %\hline       
  (y-m-d) & & &  \\   \hline
  2015-05-03&41	& 16.65$\pm$0.06	& 16.14$\pm$0.08	& 16.08$\pm$0.06 \\
  2015-06-21&90	& 17.68$\pm$0.09	& 17.16$\pm$0.09	& 16.72$\pm$0.07 \\
  2015-07-05&104& 17.97$\pm$0.07	& 17.40$\pm$0.08	& 17.36$\pm$0.08 \\
  \hline
\end{tabular}\label{tab:phot}
\end{minipage}\end{center}
\end{table}

We obtained 2 epochs of moderate-resolution optical spectra with the Bluechannel spectrograph on the Multiple Mirror Telescope (MMT) on 2015 Apr. 30 and 2015 Jun. 12. Each MMT Bluechannel observation was taken with a 1.0$\arcsec$ slit and the $1200\text{ l mm}^{-1}$ grating with a central wavelength of 6350~\AA\ and $3 \times 1200$~s exposures. The spectral range we used covers approximately 5800--7000~\AA. Standard reductions were carried out using IRAF\footnote{IRAF, the Image Reduction and Analysis Facility, is distributed by the National Optical Astronomy Observatory, which is operated by the Association of Universities for Research in Astronomy (AURA) under cooperative agreement with the National Science Foundation (NSF).} including bias subtraction, flat fielding, and optimal extraction of the spectra. Flux calibration was achieved using spectrophotometric standards observed at an airmass similar to that of each science frame, and the resulting spectra were median combined into a single 1D spectrum for each epoch.

In addition, we retrieved the spectrum used for the original spectroscopic identification of PS15si by the Public ESO Spectroscopic Survey of Transient Objects \citep[PESSTO; see][]{smartt+13,valenti+14} from WISeREP\footnote{\url{wiserep.weizmann.ac.il/}} \citep{yaron+12}; it had been obtained with the Faint Object Spectrograph and Camera (EFOSC2) on the European Southern Observatory's New Technology Telescope (ESO-NTT) on 2015 Mar. 28 \citep{atel7308}. The slit width, observed spectral range, and resolution are given in \autoref{tab:spec}.
 
\begin{table}
\begin{center}\begin{minipage}{3.3in}
      \caption{Optical Spectroscopy of PS15si}\scriptsize
\begin{tabular}{@{}cccccc}\hline\hline
  UT Date    &day\footnote{Since discovery on 2015 Mar. 23.}& Telescope/Instrument   &Slit Width & $Res.$  & Range \\ %\hline       
  (y-m-d)    &   & &    (arcsec)  & $\frac{\lambda}{\Delta \lambda}$ & (\AA)       \\   \hline
2015-03-28&5 & ESO-NTT/EFOSC2  &1.0 & 355 & 3640--9235 \\
2015-05-01&39& MMT/Bluechannel &1.0 & 4500 & 5727--7012 \\
2015-06-13&82& MMT/Bluechannel &1.0 & 4500 & 5727--7012 \\
2015-06-16&85& Keck/LRIS       &1.0 & 600 & 3200--10200 \\ 
\hline
\end{tabular}\label{tab:spec}
\end{minipage}\end{center}
\end{table}

We obtained the final spectral epoch of PS15si on 2015 Jun. 16 with the Low Resolution Imaging Spectrometer \citep[LRIS;][]{oke+95} at the Keck Observatory. We used the 1.0$\arcsec$ slit rotated to the parallactic angle to minimise the effects of atmospheric dispersion \citep[][in addition, LRIS has an atmospheric dispersion corrector and the object was at low airmass, $\sim 1.3$]{filippenko82}. In our LRIS configuration, coverage in the blue with the 600/4000 grism extends over 3200--5600~\AA. We used the 5600~\AA\ dichroic and our coverage in the red with the 400/8500 grating extends over 5600--10,200~\AA. We obtained one 200~s exposure, and reduced it using routines written specifically for LRIS in the Carnegie {\sc Python} ({\sc CarPy}) package. We performed standard reductions on the two-dimensional (2D) images including flat-fielding, correction for distortion along the slit axis, wavelength calibration with arc-lamp spectra, and cosmic ray cleaning before extracting the 1D spectrum of PS15si. We flux calibrated the extracted spectrum using a sensitivity function derived from a standard star obtained the same night in the same instrument configuration. We also used the standard-star spectrum to remove the telluric sky absorption features. 

\begin{figure*}
	\includegraphics[width=\textwidth]{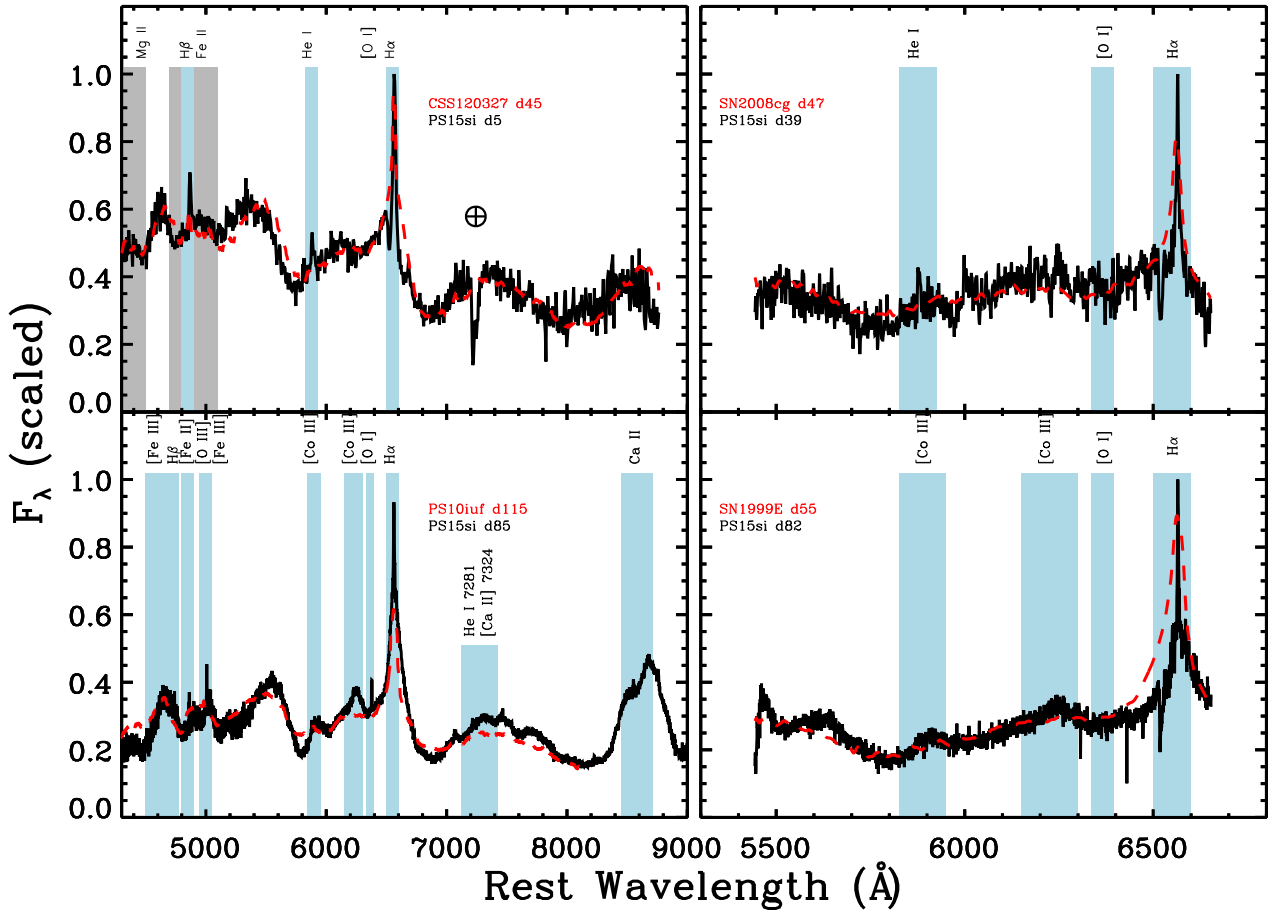}
	\caption{Four spectral epochs of PS15si with the day relative to discovery (d\#\#) of observation given in black. Overplotted at each epoch is a comparison SN~Ia/IIn spectrum in red taken from \citet{silverman+13} and corrected for redshift and Galactic reddening using the values in Tables 1 and 2 of that publication. The phase (i.e., d\#\#) for each comparison spectrum is the time in days since $R$-band maximum for CSS120327 and since discovery for SN~2011jb.  Features labeled in grey (Mg\II, Fe\II) are in absorption while features labeled in blue (H$\alpha$, H$\beta$, He\I, [O\I], [O\III], [Fe\II], [Fe\III], He\I\ $\lambda$7281/[Ca\II] $\lambda$7324 blend, Ca\II, [Co\III] $\lambda\lambda$5890, 5908,6197) are in emission.  We also indicate a telluric feature (O$_{2}$ A) in the first spectral epoch with a $\oplus$ symbol.}\label{fig:spec-comp}
\end{figure*}

The final spectra are shown in \autoref{fig:spec-comp}. The resolution at H$\alpha$ in each spectrum (in chronological order) is roughly $840, 66, 66,$ and $500~\kms$. These data have been dereddened for Milky Way Galaxy reddening assuming $E(B-V)=0.046\mags$ as reported by \citet{sf11}. We have also removed the recession velocity $v=15,747~\kms$ given the observed wavelength of narrow H$\alpha$ emission, which is consistent with the host-galaxy redshift.

\section{RESULTS}\label{sec:res}

\subsection{Photometry}\label{sec:phot-result}

To analyze our photometry of PS15si, we compare our data to light curves of other SNe~Ia in the literature, especially those of SNe~Ia/IIn after optical maximum. In \autoref{fig:phot} (left), we overplot light curves from an overluminous SN~1991T-like template \citep{stern+04} and SN~2005gj as a SN~Ia/IIn template \citep{prieto+07} along with our $R$-band observations and the discovery $w$-band magnitude of PS15si. The light curve of SN~2005gj is fit to our data of PS15si assuming that both light curves follow the same shape, and that PS15si is $0.56\mags$ fainter than SN~2005gj (i.e., shifted in magnitude without any stretch). This brightness allows us to match the shape of the SN~2005gj light curve to prediscovery constraints detailed in \autoref{sec:date}.

As we demonstrate in \autoref{fig:phot} (right), the $R$-band magnitude of PS15si appears to be in a phase of linear decay and at a rate of $0.011~\text{mag day}^{-1}$, although later epochs in the $B$ and $V$ bands may be decaying more slowly or even rebrightening. We explore this behaviour in \autoref{sec:late}.

\subsection{Spectroscopy}\label{sec:spectros}

\subsubsection{Comparison to Other SNe~Ia/IIn and Spectral Line Identification}\label{sec:comp-line}

In \autoref{fig:spec-comp} we compare spectra from each available epoch of PS15si to spectra of known SNe~Ia/IIn. We note several features in absorption (grey) and in emission (blue) present in each spectral epoch. The comparison spectra were sampled from all of the SNe~Ia/IIn spectra presented in \citet{silverman+13}. For each epoch, the time (in days) since discovery is indicated for PS15si and the comparison spectra, including CSS120327:110520–015205 \citep[hereafter, CSS120327; identified as a SN~Ia/IIn in][]{drake+12}, SN~2008cg \citep[identified as a SN~Ia/IIn in][]{filippenko+08}, SN~1999E \citep[identified as a SN~Ia/IIn in][]{deng+04}, and PTF10iuf \citep[identified as a SN~Ia/IIn in][]{silverman+13}. The PS15si spectra are scaled such that $F_{\text{H}\alpha} = 1$, and the comparison spectra are corrected for redshift and Galactic reddening using values from \citet{silverman+13} (Tables 1, 2, 4, and 5 therein) and scaled to minimise the root-mean square (RMS) of the difference spectrum except for H$\alpha$ (i.e., 6500--6600~\AA).

The correspondence between H$\beta$, He\I\ $\lambda$5876, Mg\II\ around 4400~\AA, and Fe\II\ 4700--5100~\AA\ in the first spectral epoch of PS15si and the spectrum of CSS120327 indicates the similarities between these objects. In our two moderate resolution epochs (\autoref{fig:spec-comp}, right panels), H$\alpha$ is narrower in the PS15si spectra than in the SN~2008cg and SN~1999E spectra.  This discrepancy could be due to the difference in spectral resolution of the PS15si epochs and comparison spectra, especially as the comparison spectra appear to peak at a lower flux density relative to the continuum level.  If the discrepancy is real, however, it could also be due to the relative difference in epochs between PS15si and the comparison spectra or an intrinsic difference in the strength of post-shock CSM.  That is, the intermediate-width component of H$\alpha$, which begins to appear 82 days after discovery in PS15si, is significantly weaker in PS15si than in the comparison spectra.  This feature presumably arises from the shell of post-shock CSM entrained in the ejecta \citep[][]{chevalier+94,zhang+12}, and therefore indicates the relative epoch of PS15si and the comparison spectra assuming similar CSM profiles.

In the final epoch, there are additional differences in the comparison to PTF10iuf.  Broad emission features are generally  well-matched to PTF10iuf, although the continuum level appears slightly higher in the comparison spectrum.  Emission around 5900~\AA\ and especially 6200~\AA\ is also stronger in PS15si than in PTF10iuf.  These features are of particular interest because they correspond to known Co features in the nebular spectra of SNe~Ia (i.e., [Co\III] $\lambda\lambda$5890, 5908 and [Co\III] $\lambda$6197 as described in \citet{bowers+97,liu+97}).  In addition to [Co\III] $\lambda$6578 (which is obscured by H$\alpha$ in SNe~Ia/IIn), these Co lines are generally the strongest Co features in optical spectra of SNe~Ia. We interpret this line identification as independent confirmation that PS15si is likely a SN~Ia/IIn and results from a thermonuclear SN; SNe~IIn generally do not have strong Co emission in their late-time spectra but are instead dominated by narrow lines of H, O, and He as well as a quasicontinuum of forbidden and permitted Fe lines \citep[as in SNe~1988Z, 1998S, 2006gy][]{turatto+93,mauerhan+12,kawabata+09}. The presence of strong Co emission in addition to the quasicontinuum of Fe lines at bluer wavelengths strongly suggests that the underlying SN is of Type Ia.

One strong similarity between PS15si and other SNe~Ia/IIn in the final epoch is the presence of narrow [O\III] $\lambda$5007 emission, which is also accompanied by narrow [O\I] $\lambda$6364. PS15si represents the first time [O\I] emission has been identified from a SN~Ia/IIn. The emission from these features cannot be due to host galaxy contamination given the lack of any such emission in the first spectral epoch. Therefore, we infer that the [O\I] and [O\III] emission arise from either the SN ejecta or the surrounding medium. In spectra from SN~2005gj, \citet{aldering+06} found that [O\III] around $71\days$ past explosion exhibited an inverted P Cygni profile with emission around $-100~\kms$ and absorption around $+450~\kms$. Our spectral resolution in the final epoch ($\sim 500~\kms$) cannot resolve any P Cygni feature in [O\III] or [O\I] if it exists. This line may also be obscured by strong Fe emission at similar wavelengths\footnote{We mark [Fe\II] $\lambda$4890 and [Fe\III] $\lambda$5011 in \autoref{fig:spec-comp} for reference.}.

\begin{figure}
\includegraphics[width=0.475\textwidth]{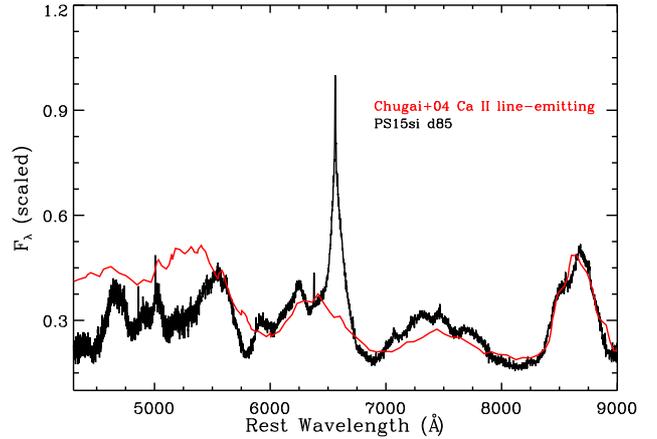}
\caption{Final spectral epoch of PS15si (2015 Jun. 16) in black. In red, we show a comparison, synthetic spectrum taken from \citet[][Figure 10(c) therein]{chugai+04}.}\label{fig:caII}
\end{figure}

The Ca\II\ IR triplet feature appears to be one of the dominant spectral lines in the final spectral epoch of PS15si. For SN~2002ic, \citet{chugai+04} argued that Ca\II\ emission around 8500~\AA\ points to a Ca-rich layer in the shocked SN ejecta arising from an outer layer of Fe-poor material. In \autoref{fig:caII}, we show that their model of a Ca\II\ line-emitting shell closely matches the profile observed toward PS15si in the latest epoch. However, that quasicontinuum, Fe-dominated model around 4300--6000~\AA\ is poorly matched to PS15si. Strong Ca\II\ emission is a sign that incomplete burning up to Fe-peak products may accompany some SNe~Ia/IIn \citep{marion+03}, which can occur in thermonuclear SNe in general\footnote{However, note that \citet{mazzali+05} discuss abundance enhancement versus density enhancement in combination with SN-CSM interaction to explain Ca II IR absorption features in some normal SNe~Ia.}. This prominent, broad Ca\II\ profile is common for SNe~Ia/IIn; indeed, it appears in virtually all SNe of this type with spectra available $\sim70-400\days$ after optical maximum \citep[e.g., SNe 1999E, 2002ic, 2008J, 2011jb, PTF11kx, 2012ca in][]{rigon+03,wang+04,silverman+13,silverman+13b,fox+15}.  This feature contrasts with spectra of ``normal'' SNe~Ia in which Ca\II\ is generally weak during the nebular phase \citep{bowers+97}, although it has been observed at very late times \citep[e.g., $> 500\days$ in SN~2011fe as in][]{graham+15}.

\subsubsection{Fitting PS15si to Spectral Templates with Added Blackbody Emission}\label{sec:fitting}

Apart from strong, narrow H$\alpha$ emission, the presence of luminous, optically thick continuum is one of the defining characteristics of SNe~Ia/IIn that led to their interpretation as CSM interaction events \citep{hamuy+03,aldering+06,leloudas+15}. Strong continuum emission tends to ``dilute'' the SN~Ia spectrum, as emission and absorption features lose contrast relative to the continuum level. This is seen in core-collapse SNe~IIn with strong CSM interaction as well \citep[e.g., SN~2006gy as in][]{smith+10}. Assuming the continuum emission is dominated by thermal emission from CSM interaction, we can constrain the temperature of the ejecta-CSM region and subtract this component to examine the underlying SN~Ia-like features. This analysis is based on the assumption that the SN itself is typical of SNe without signatures of CSM interaction at varying epochs. In the case of PS15si, we simultaneously fit a blackbody continuum and SNe templates to the spectrum at each epoch. These templates include SN~1994D-like (SN~Ia-norm; 0--90 days after explosion), SN~1991T-like (overluminous SN~Ia; 0--93 days), SN~1991bg-like (underluminous SN~Ia; 0--113 days) and SN~1999ex-like (SN~Ib/c; 0--85 days) spectra \citep{nugent+02,hamuy+02,stern+04}\footnote{\url{c3.lbl.gov/nugent/nugent_templates.html}}. While we have established that PS15si is spectroscopically similar to other SNe~Ia/IIn and has spectral features typical of SNe~Ia, we use a SN~Ib/c template in our analysis as an additional check on the hypothesis that PS15si is more similar to SNe~Ia than other types.

The template spectra ($P_{0}(\lambda)$) we obtained were in arbitrary units, and we normalized them by taking $P_{\lambda} = \frac{P_{0}(\lambda)}{P_{0}(\mathrm 5000~\mathrm{\AA})}$. For each PS15si spectrum $F_{\lambda}$, we simultaneously fit a template spectrum $P_{\lambda}$ and a blackbody continuum spectrum normalized to unity at its peak ($\lambda_{peak} = \frac{h c}{4.965 k_{B} T}$) such that $B_{\lambda}(T) = \left(\frac{h c}{1.842 k_{B} T \lambda}\right)^{5} \left(\text{exp}\left(\frac{h c}{\lambda k_{B} T}\right)-1\right)^{-1}$ with variable temperature $T$ and the relative weights of the template and blackbody $C$ and $D$, respectively. Thus, we searched for parameters, $C,D,T$ to minimise the RMS of the difference spectrum

\begin{equation}
		Z_{\lambda} = F_{\lambda} - C \times P_{\lambda} - D \times B_{\lambda}(T) 
\end{equation}

\noindent Our results from spectral fits at all four epochs of PS15si are presented in red, blue, orange, and green for each template in \autoref{fig:spec-fit}.  The template spectra are labeled with the epoch in days relative to explosion along with the temperature of the blackbody continuum added to the fit.  Below the best-fit spectrum from each template, we plot the residuals for all four templates (i.e., $Z_{\lambda}$).  The RMS of $Z_{\lambda}$ for each spectral fit is given in \autoref{tab:rlap}.  In calculating the RMS, we ignore wavelength ranges that may be fit to features not present in the template spectra, such as H$\alpha$ (6400--6700~\AA) as well as the telluric feature in the first epoch (7200--7300~\AA).  We consider other features to be too weak to skew our fitting routine significantly and thus we include, for example, the wavelength range around H$\beta$, which also contains Fe\II\ features.  In this way, the statistic that we use to fit template spectra to epochs of PS15si, the RMS of $Z_{\lambda}$, is useful primarily to compare different templates and fitting parameters within a single epoch and not as an overall ``quality of fit'' parameter.  In \autoref{sec:spec-qual}, we assess the overall quality of our spectral fitting and alternative statistics in order to determine which spectral template best matches PS15si. 

We report the best-fit values $C,D,$ and $T$ for each epoch of PS15si and set of template spectra (SN~Ia 1991bg-like, SN~Ia 1991T-like, SN~Ib/c 1999ex-like, and SN~Ia-norm 1994D-like) in \autoref{tab:fit-values}.  We also include the ratio of the weighted template spectrum to blackbody emission ($C \times P_{\lambda}$ to $D \times B_{\lambda}(T)$) in the $V$ band (i.e., 5028--5868~\AA), which we refer to as $f_{V}$ \citep[see][]{leloudas+15}.  For each of the three fitting parameters, we approximate an uncertainty by fixing two parameters at their best-fit values and varying the third parameter until the RMS of $Z_{\lambda}$ (given in \autoref{tab:rlap}) increases by a factor of $2$.

\begin{table}
\begin{center}\begin{minipage}{4in}
      \caption[Best-Fit Parameters for Spectral Decomposition of PS15si]{Best-Fit Parameters $C$,$D$,$T$, and $f_{V}$ used to generate spectra in \autoref{fig:spec-fit}.}\scriptsize
\begin{tabular}{cccccc}\hline\hline
  Template & Par.\footnote{$C$ and $D$ in units of $10^{-17}~\text{erg s}^{-1}~\text{cm}^{-2}$~\AA$^{-1}$, $T$ in units of K.} & 2015-03-28 & 2015-05-01 & 2015-06-13 & 2015-06-16 \\ \hline       
  91bg-like & $C$		& $0.9\substack{+3.9\\-0.9}$ & $1.4\substack{+1.7\\-1.4}$ & $3.5\substack{+3.7\\-3.5}$  & $5.9\substack{+8.0\\-5.9}$  \vspace{0.02in}\\
  	 		  & $D$		& $27\substack{+7.8\\-7.7}$  & $7.5\substack{+2.3\\-2.3}$ & $12.5\substack{+4.5\\-4.6}$ & $10.8\substack{+6.8\\-6.6}$ \vspace{0.02in}\\
  			  & $T$		& $6400\substack{+2300\\-1600}$ & $3200\substack{+18000\\-1100}$ & $2700\substack{+1500\\-1100}$ & $4500\substack{+7700\\-2100}$ \vspace{0.02in}\\
			  & $f_{V}$ & $0.07$ & $0.33$ & $0.48$ & $0.54$ \vspace{0.1in}\\
  91T-like & $C$		& $6.1\substack{+7.0\\-6.1}$ & $1.0\substack{+1.3\\-1.0}$ & $2.8\substack{+2.5\\-2.7}$ & $11.1\substack{+6.5\\-6.2}$ \vspace{0.02in}\\
  	  		 & $D$	    & $21.4\substack{+6.3\\-6.4}$ & $7.1\substack{+2.4\\-2.4}$ & $11.0\substack{+4.6\\-4.7}$ & $12.1\substack{+14.4\\-12.1}$ \vspace{0.02in}\\
			 & $T$  	& $5600\substack{+2000\\-1700}$ & $3400\substack{+26000\\-1300}$ & $2800\substack{+19000\\-1300}$ & $2100\substack{+1700\\-2100}$ \vspace{0.02in}\\
			 & $f_{V}$	&  $0.32$ & $0.31$ & $0.62$ & $12.5$ \vspace{0.1in}\\
 99ex-like & $C$ 		& $4.8\substack{+7.4\\-4.8}$ & $1.2\substack{+1.8\\-1.2}$ & $0.1\substack{+3.0\\-0.1}$ & $8.3\substack{+5.1\\-5.2}$ \vspace{0.02in}\\
  			 & $D$  	& $24.9\substack{+7.7\\-7.6}$ & $6.8\substack{+2.3\\-2.2}$ & $13.7\substack{+4.0\\-4.4}$ & $7.4\substack{+5.6\\-5.7}$ \vspace{0.02in}\\
			 & $T$  	& $6500\substack{+2700\\-1700}$ & $3700\substack{+42000\\-1400}$ & $3500\substack{+17000\\-1400}$ & $4000\substack{+5400\\-2400}$ \vspace{0.02in}\\
			 & $f_{V}$  & $5.4$ & $6.6$ & $141$ & $1.2$ \vspace{0.1in}\\
  94D-like & $C$ 		& $7.8\substack{+7.1\\-7.1}$ &  $1.1\substack{+1.9\\-1.1}$ & $3.8\substack{+3.1\\-3.4}$ & $4.7\substack{+7.6\\-4.7}$ \vspace{0.02in}\\
  	  		 & $D$  	& $19.7\substack{+6.3\\-6.2}$ & $7.4\substack{+2.2\\-2.3}$ & $10.8\substack{+4.0\\-4.1}$ & $11.3\substack{+7.0\\-6.9}$ \vspace{0.02in}\\
			 & $T$		& $5400\substack{+2200\\-1700}$ & $3500\substack{+42000\\-1200}$ & $3000\substack{+15000\\-1300}$ & $4500\substack{+7700\\-2200}$ \vspace{0.02in}\\
			 & $f_{V}$  & $0.44$ & $0.21$ & $0.53$ & $0.43$ \\
  \hline
\end{tabular}\label{tab:fit-values}
\end{minipage}\end{center}
\end{table}

\subsubsection{Quality of Spectral Fitting and the SN~Ia Spectrum Underlying PS15si}\label{sec:spec-qual}

\begin{figure*}
\includegraphics[width=\textwidth]{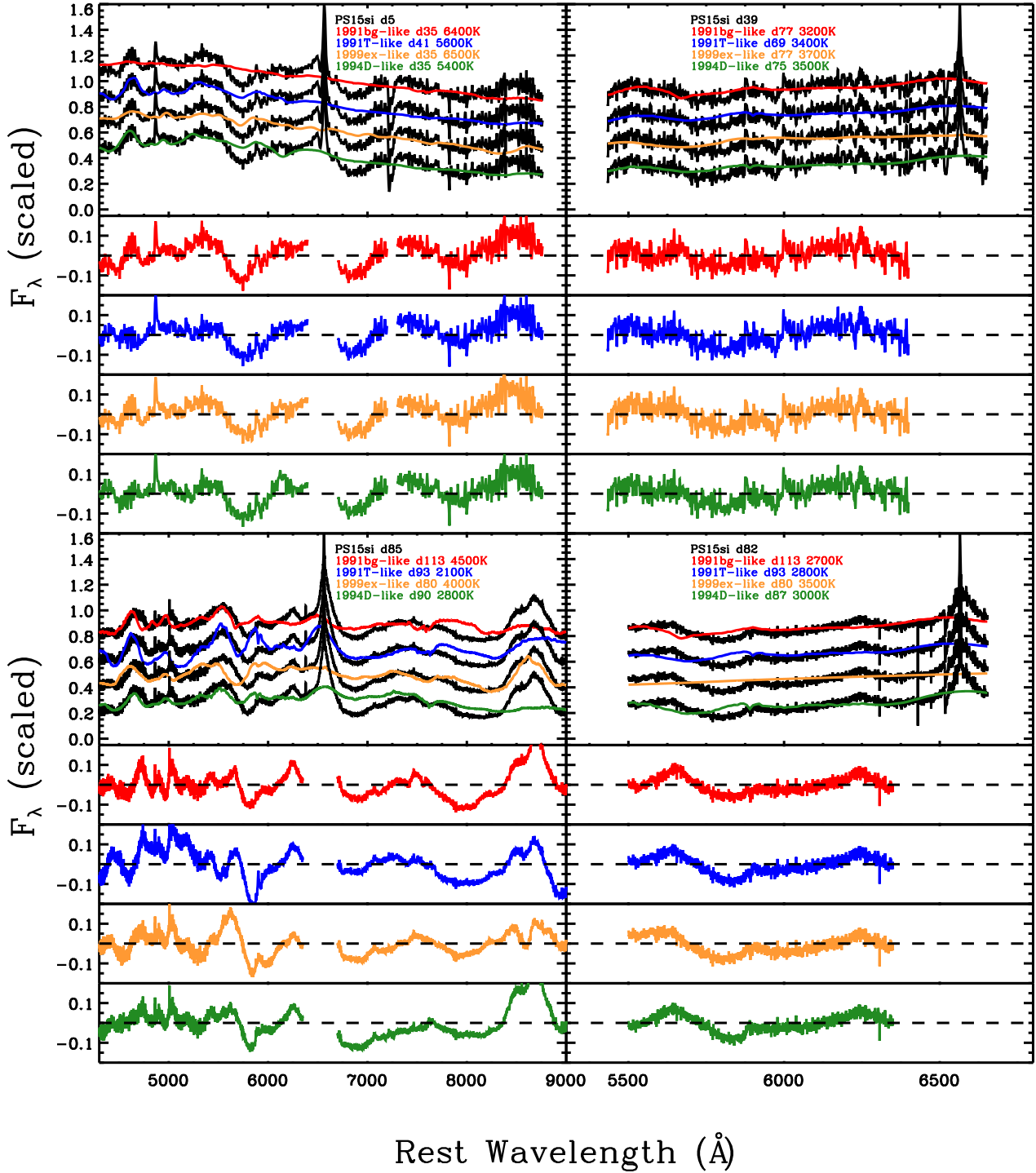}
\caption{Four spectral epochs of PS15si with the day relative to discovery (d\#\#) given in black. We also plot comparison SN~Ia 1991bg-like templates (red), SN~Ia 1991T-like templates (blue), SN~Ib/c 1999ex-like templates (orange), and SN~Ia-norm 1994D-like templates (green) plus blackbody continuum. These template spectra are derived from those available in \citet[SN~1991bg-like and 1994D-like, also \url{c3.lbl.gov/nugent/nugent_templates.html}]{nugent+02}, \citet[SN~1991T-like]{stern+04}, and \citet[1999ex-like]{hamuy+02}. We label the template spectrum with the epoch (d\#\#) in days relative to explosion as well as the temperature of the added blackbody continuum. We also plot the residuals for each comparison spectrum below the stacked spectra.}\label{fig:spec-fit}
\end{figure*}

As \autoref{fig:spec-fit} and \autoref{tab:fit-values} demonstrate, the statistical leverage in identifying PS15si comes mostly from the first and last epochs given the much wider spectral range.  Indeed, the spectral identification, including type and epoch, as well as the intensity and temperature of blackbody emission in the other two (high-resolution) epochs is highly uncertain.  While we include these spectral epochs in most of our analysis, they do not contribute significantly to our conclusions regarding spectral fitting.

The analysis described above is similar to the Monte Carlo simulation used by \citet{leloudas+15} to fit SNe~Ia/IIn spectra.  Similarities between these analyses are that the template spectrum $P_{\lambda}$ spans a range of types including SNe~Ia-norm, SN~1991T-like, SN~1991bg-like, and SNe~Ib/c, the variables $D$ and $T$ account for the blackbody radius and temperature, respectively, and we fix the spectral resolution of the template and blackbody to that of the input PS15si spectrum.  Unlike the \citet{leloudas+15} analysis, we do not explicitly fit the line spectrum associated with CSM interaction or artificially inject noise into our comparison spectra, we fix the reddening to the value noted above for PS15si (i.e., we use a dereddened PS15si spectrum as input), and we do not use human classifiers but rather rely solely on statistical analysis to identify the best-fit spectra.

A key parameter in the \citet{leloudas+15} analysis is $f_{V}$, which we describe in \autoref{sec:fitting} and include for each PS15si epoch and template spectrum in \autoref{tab:fit-values}.  One of the central arguments in \citet{leloudas+15} is that, when $f_{V}$ is large (e.g., $>1$), dilution from the CSM emission is low enough that spectral identification is more accurate. Given that CSM emission is dominant at late times after explosion, this analysis would suggest that earlier spectra are more reliable when identifying the underlying SN.

Our analysis reveals some curious trends in matching PS15si to SN templates of varying types.  In general, the SN~Ib/c 1999ex-like spectra yield poor fits to PS15si with the SN emission dominant over the CSM emission ($f_{V} > 1$) in all epochs.  This trend and the large uncertainties on fitting parameters for the SN~Ib/c template suggest this spectrum is unlikely to describe the underlying SN emission from PS15si.

In addition to our analysis in \autoref{sec:comp-line}, from the fact that PS15si does not exhibit a clearly broad H$\alpha$ component from emission in its ejecta and the poor fit between PS15si and SN~Ib/c templates, we infer that PS15si must be a better match to SN~Ia templates.  This appears to be the case, as in each spectral epoch the SN~Ia templates yield lower RMS values and the best fits to PS15si overall.  While there is no clear preference in the RMS values for any particular template, the $f_{V}$ values and uncertainties on the fitting parameters suggest that the SN~1991bg-like template is not physically plausible. For example, in the first spectral epoch when emission from the underlying SN should be most dominant, $f_{V}=0.07$ for the SN~1991bg-like template makes it clear that the template is fitted poorly to the PS15si spectrum.  The RMS value for this spectral fit supports the hypothesis that the SN~1991bg-like template is the poorest fit to PS15si in this epoch. In the final epoch, however, PS15si is better fit to the SN~1991bg-like template, although that may be due to the relative lack of continuum emission in underluminous SN~1991bg-like SNe~Ia or else the fact that the available epochs of the SN~1991T template (0--93 days) do not cover as large a range as the SN~1991bg templates (0--113 days). As we discuss below for the SN~1991T-like template, it is difficult to decompose a SNe~Ia/IIn into its constituent SN and CSM spectrum given the degeneracy between these components.

One might expect that if SN~1991bg-like SNe~Ia are a better match to SNe~Ia/IIn than SN~1991T-like SNe~Ia (or vice versa), then SNe~Ia-norm should fall between the two in terms of quality of fit.  This trend appears to be the case in both the first and last epochs when PS15si matches the 1994D-like template about as well as the SN~1991T-like and SN~1991bg-like templates, respectively.  The value of $f_{V}$ is comparable to what \citet{leloudas+15} find for post-maximum SNe~Ia/IIn templates, although it barely decreases between the best-fit values for the first and last epochs of the 1994D-like template, contrary to the observation that SNe~Ia fade much more rapidly than CSM emission.  Again, there are problems in that the decomposition does not fully recover the relative contribution of these components in a physically consistent manner.

Based on the uncertainties and the self-consistent evolution of the temperature across all four epochs, SN~1991T appears well-fit to PS15si.  However, in the final epoch there are some differences on the blue end of the spectrum that suggest there is extra CSM emission diluting the continuum between emission features as well as much weaker emission around 5800--6000~\AA, which in SN~1991T was identified in nebular spectra as [Co\III] \citep{schmidt+94}.  Overall, SN~1991T is the poorest fit to PS15si in this epoch and the fit to SN~1991T involves an extremely large value of $f_{V}$ ($=12.5$) for this late epoch, although both of these effects might be explained by the poor approximations involved in decomposing a SN~Ia/IIn spectrum into a SN~Ia and CSM emission.  For example, SN~1991T-like SN~Ia spectra are associated with overluminous and hot SNe with significant $^{56}$Ni production and whose spectra exhibit high-ionisation lines \citep{filippenko+92a}.  Perhaps the high luminosity that is associated with $^{56}$Ni production in SN~1991T-like SNe~Ia instead comes almost entirely from CSM interaction in SN~Ia/IIn, which would explain the poor fit to [Co\III] in the final epoch of PS15si as well as the poor fit to continuum emission between spectral lines and the unusually large $f_{V}$ ratio.  We discuss further ramifications of this idea in \autoref{sec:behaviour}.

\begin{table}
\begin{center}\begin{minipage}{3.3in}
      \caption{RMS of Residuals for Each Template in \autoref{fig:spec-fit}}\scriptsize
\begin{tabular}{@{}ccccc}\hline\hline
  UT Date & 91bg-like & 91T-like & 99ex-like & 94D-like \\ %\hline       
  (y-m-d) & & & &    \\   \hline
  2015-03-28 & 0.067 & 0.062 & 0.065 & 0.063 \\
  2015-05-01 & 0.048 & 0.051 & 0.052 & 0.050 \\
  2015-06-13 & 0.039 & 0.043 & 0.045 & 0.041 \\
  2015-06-16 & 0.049 & 0.081 & 0.065 & 0.049 \\
  \hline
\end{tabular}\label{tab:rlap}
\end{minipage}\end{center}
\end{table}

\section{DISCUSSION}\label{sec:disc}

\subsection{PS15si Explosion Date and Maximum $R$-Band Magnitude}\label{sec:date} 

Comparison between our photometry and light curves from other SNe~Ia/IIn provides a method to infer the phase of PS15si. We acknowledge that this analysis assumes that the shape of the PS15si light curve near maximum, which is not well-constrained by our data, follows that of other SNe~Ia/IIn. Recent analysis comparing several SNe~Ia/IIn \citep[SNe 2005gj, PTF11kx, 2002ic, 1999E, 1997cy in][and Figure 4 therein]{inserra+16} found that light curves from most SNe~Ia/IIn have a similar shape and bolometric luminosity from early times to as late as $\sim 200\days$ after optical maximum except for the unusual case of PTF11kx, whose bolometric luminosity fades much more quickly and is very similar to SN~1991T.  For the majority of SNe~Ia/IIn, variations in CSM profiles for each object may cause bolometric light curves to diverge significantly after $\sim200\days$. SN~Ia/IIn absolute magnitudes also span at least $1.8\mags$ in the $r$ band at maximum brightness \citep{silverman+13}, so calibrating a light curve to these objects will have similarly large systematic uncertainties.

In \autoref{fig:phot}, we compare the available $R$-band light curve of PS15si to the exponential decay (i.e., in flux versus time) phase of SN~2005gj \citep{prieto+07}. Estimating the $R$-band maximum brightness of PS15si involves a degeneracy between the maximum brightness and the epoch of observation. For example, we fit the SN~2005gj light curve to PS15si assuming these objects have the same absolute magnitude in $R$, but PS15si could have reached optical maximum earlier with a brighter maximum. From our $R$-band photometry in \autoref{fig:phot}, SN~2005gj appears to be declining at a rate of $0.011~\text{mag day}^{-1}$, which is close to the decay rate observed in our photometry of PS15si. This rate is also significantly slower than the $0.025~\text{mag day}^{-1}$ observed for $^{56}$Co decay in normal SNe~Ia during the epoch $30-90~\text{days}$ after explosion \citep[see, e.g.,][]{phillips+99,forster+13}, although it is consistent with the decline rates observed in the $R$ band for other interaction-powered SN light curves \citep[e.g., SNe 2005cp and 2005db as in][]{kiewe+10}.

Assuming the PS15si light curve follows the same general shape as SN~2005gj in $R$ and PS15si had a maximum $R$-band absolute magnitude of $M_{R,{\rm max}}$, we find that the time (in days since discovery) of optical maximum for PS15si was

\begin{equation}
		t_{\rm max} = -18+\left(\frac{M_{R,{\rm max}}+20.36}{0.021}\right).
\end{equation}

\noindent Therefore, an $R$-band maximum brightness equivalent to that of SN~2005gj ($-20.36\mags$) implies that PS15si reached maximum light at $t = -18\days$ or on 2015 Mar. 5. This phase roughly agrees with that of \citet{atel7308}, in which the spectral fits to SN~IIn models predicted that PS15si was $> 20\days$ after optical maximum on 2015 Mar. 28.

However, nondetections were reported toward PS15si in the  $w$ band to deep limits of $m_{w} = 22.0\mags$ (i.e., $M_{w} = -14.8\mags$) on 23 and 24 Feb. 2015\footnote{These data are available from the Pan-STARRS Survey for Transients at \url{star.pst.qub.ac.uk/ps1threepi/} as described in \citet{huber+15}}, $27\days$ before discovery, implying that the SN had not yet exploded. Assuming PS15si needed between 18 and 45 days to rise to maximum \citep[e.g., the minimum and maximum rise times found for SNe~Ia/IIn by][]{silverman+13}, then comparison to SN~2005gj via Equation (2) implies that PS15si reached $R$-band maximum approximately 18 to 45 days after the last nondetection and with a magnitude in the range $-20.2\mags < M_{R} < -19.6\mags$. \autoref{fig:phot} demonstrates these constraints and shows a comparison of PS15si to the SN~2005gj light curve. In this scenario, PS15si was discovered on 23 Mar. 2015 (the $w$-band measurement in \autoref{fig:phot}), $8\days$ before $R$-band maximum, and had a maximum brightness of $M_{R} = -19.68$.  In this model, we assume that PS15si took $\sim34\days$ to rise in the $R$ band (i.e., it exploded $26\days$ before discovery), which matches the behaviour of SN~2005gj \citep{prieto+07}.

\subsection{Luminosity of the CSM and Underlying SN}\label{sec:behaviour}

For CSM-interacting SNe at late times and for a high wind-density parameter $w = 4 \pi r^{2} \rho > 10^{16}~\text{g cm}^{-1}$, the overall luminosity of the radiative forward shock wave with velocity $v_{s}$ will be $L \propto w v_{s}^{3}$ \citep{cy04}; this is the rate at which the shock sweeps up mass ($4 \pi r^{2} \rho v_{s}$) times the energy per unit mass in the shock ($v_{s}^{2}$).

Following the treatment presented in \citet{ofek+14}, we assume that $\rho = b r^{-k}$, which implies $L \propto r^{2-k} v_{s}^{3} \propto t^{-\frac{2k-7}{k-4}}$, where we have implicitly assumed that $\rho_{\rm ej} \propto t^{-3} (\frac{r}{t})^{-m}$, and $m = 4$, which holds for a momentum-conserving ``snowplough'' SN\footnote{\citet{ofek+14} find $L=L_{0} t^{\alpha}$, where $\alpha=\frac{(2-k)(m-3)+3(k-3)}{m-k} = -\frac{2k-7}{k-4}$, $r~\propto~t^{(m-3)/(m-k)} = t^{-1/(k-4)}$ for $m=4$.} \citep{svirski+12}.

Assuming the CSM component of the $R$-band data in \autoref{tab:phot} and \autoref{fig:phot} follows a power-law with $L_{R} \propto t^{a}$, i.e., $M_{R} \approx -2.5 a \log t$ (with $t$ in days since discovery), we subtract the $R$-band luminosity from a SN~1991T-like template from our observed $R$-band magnitudes to determine the contribution to the luminosity from CSM emission (i.e., $L = L_{CSM} + L_{SN}$). We find that the residual magnitude is well-fit by $M_{R} \approx 1.82 \log t$, which implies $L_{R} \propto t^{-0.727 t}$. This trend in luminosity implies $k = 3.5$. The inferred power-law is relatively insensitive to the explosion time. For example, if we model the $R$-band luminosity versus time since explosion (by adding $26$~days according to our analysis in \autoref{sec:date}), then we find $M_{R} \approx 2.56 \log t$, which implies $L_{R} \propto t^{-1.02 t}$ and $k=3.7$. Clearly, the density profile implied by our photometry and this model is very steep. Indeed, the self-similar CSM treatment of \citet{chevalier82} breaks down for $k > 3$, although similar results have been reported in the much later ($t\approx300\days$) optical light curve of SN~2010jl \citep{ofek+14}.  

From our spectral fitting to SN~1991T-like templates, we find that a thermal continuum component fit to the spectrum of PS15si at our first and last epochs implies best fits with $T = 5600~\text{K}$ and $2100~\text{K}$, although as we discuss below, the latter temperature may be an underestimate.  The second and third (higher-resolution) epochs have highly uncertain temperatures, which we do not include in this analysis.  Assuming the SN has expanded to $\sim 10^{16}~\text{cm}$ by $86\days$ after discovery (i.e., $100-120\days$ after explosion given the model in \autoref{sec:date} and with an expansion velocity of $10^{4}~\kms$), a spherical shell emitting at $2100~\text{K}$ will have a luminosity of about $1.4\times10^{42}~\text{erg s}^{-1}$ or $M \approx -16.7$ mag. This value is far too small to account for the luminosity observed during the final epoch ($M \approx -19$) when emission from the CSM should dominate the SN. There is some inconsistency in the overall shape of the spectrum at late times, especially as features from the SN are easily visible at a time when the CSM should outshine the SN emission by a few magnitudes \citep[e.g., \autoref{fig:phot} herein, and Figure 7 in][]{prieto+07}.

However, ultraviolet/X-ray light from the reverse shock may contribute significantly to emission from the ejecta, producing a quasicontinuum with distinct spectral features. We have found that PS15si can be fit by overluminous SN~1991T-like SNe~Ia spectra. This underlying SN type is consistent with other SNe~Ia/IIn, but there are some remaining inconsistencies in terms of the continuum level and the relative strength of the [Co\III] feature in the final epoch. Perhaps some of the continuum emission that comes from CSM interaction in SNe~Ia/IIn is misinterpreted as the hot, luminous continuum emission observed in SN~1991T-like SNe~Ia. At the same time, some of the light from the CSM interaction in SNe~Ia/IIn could be reprocessed by the ejecta, changing the ionisation state near the interaction region and heating new layers of ejecta as the SN evolves. This hypothesis would also explain why the spectral continuum of some SNe~Ia/IIn appears to evolve from thermal to nonthermal, as in SN~2002ic \citep{chugai+04} and SN~2005gj \citep{aldering+06}. In the former, the continuum emission was well-fit around 244 days after optical maximum by a quasicontinuum model composed of cool shocked ejecta emitting in lines of Fe-peak elements. Varying the mixture of elements in models of the line-emitting shocked ejecta layer may simultaneously reproduce the observed continuum level, SN~Ia-like absorption features, and anomalous line strengths such as the Ca\II\ emission we noted in \autoref{sec:spectros}.

One consequence of this hypothesis would be that, with an added source of ultraviolet/X-ray emission, we would expect to observe an increased fraction of high-ionisation lines in the ejecta.  Indeed, the correlation between most SNe~Ia/IIn and SN~1991T-like spectra may largely be due to the added radiation from the CSM interaction, as SN~1991T exhibited a blue spectrum dominated by Fe\III\ absorption around maximum \citep{filippenko+92a,ruiz-lapuente+92} and [Fe\III] and [Co\III] emission dominant in its nebular spectrum \citep{schmidt+94}.

This hypothesis may explain the spectral morphology of SNe~Ia/IIn, although the difficulty in identifying the characteristics of the underlying SN remains. For example, one might expect that the underlying SN~Ia spectrum should be correlated with the amount of radiation produced by the CSM interaction and thus the peak luminosity of the SN.  Studies suggest that SNe~Ia/IIn with relatively low and high peak luminosities \citep[e.g., SN~2008J and PTF10iuf in][]{taddia+12,leloudas+15} have been associated with SN~1991T-like spectra.  Are there added effects due to the distribution of the CSM, viewing angle, or dilution of the underlying spectrum from the CSM quasicontinuum?  In addition, the outer layers of the SN ejecta will receive most of the added radiation from the CSM interaction, and the composition of the visible ejecta layers is largely dependent on the underlying thermonuclear explosion.  While it is worth speculating on these questions, it may be impossible to disentangle the relative importance of these effects without detailed hydrodynamic and radiative-transfer models of CSM-interacting SNe.

\subsection{Late-Time Rebrightening and Spectral Fitting of SNe Ia/IIn}\label{sec:late}

PS15si appeared to be rebrightening in the $B$ band (increase of $0.31\mags$ with $1.6\sigma$ significance) and the $V$ band (increase of $0.32\mags$ with $2.3\sigma$ significance). This trend was observed in the last four epochs of our optical photometry (\autoref{tab:phot}) in $B$ and $V$, between $85$ and $90\days$ after discovery. Either the rebrightening was not as apparent in NIR observations or the trend occurred over a relatively short period of time and the SN was no longer increasing in brightness by $104\days$ after discovery. Our final spectral epoch corresponds to the beginning of this trend in the photometry.

The flux from a CSM-interacting SN light curve may look more like a power law at late times as opposed to exponential decay. This may cause the slope of the decay curve to change where the light curve appears to be leveling off as the CSM cools radiatively as in \autoref{sec:behaviour}. This hypothesis does not fully explain the behaviour in $V$, however, where the light curve is systematically brightening over the final three epochs by as much as $\sim 30\%$ in luminosity ($0.32 \pm 0.13\mags$).

This behaviour supports the conjecture that the CSM density profile around SNe~Ia/IIn can be dense and clumpy, perhaps with steep density gradients associated with these clumps \citep[e.g., as suggested by spectropolarimetry of H$\alpha$ toward SN~2002ic;][]{wang+04}. Several other SNe~Ia/IIn, such as SN~1997cy \citep{inserra+16}, have exhibited variations in their bolometric light curves over short timescales. This behaviour contrasts with rebrightening seen in the mid-IR toward SN~2005gj \citep{ff13} where the increase in luminosity was observed predominantly at longer wavelengths ($3.6-5~\mu\text{m}$), over longer timescales ($\sim600\days$), and starting at least a year after the SN was discovered. In this latter case, rebrightening was thought to be due to reprocessed light from a dust shell surrounding the SN. For PS15si, interaction between the SN ejecta and clumps or a thin shell in the CSM can account for the short timescale over which the rebrightening occurred. As with core-collapse SNe~IIn (e.g., SNe 2001em and 2006gy as in \citet{schinzel+09} and \citet{smith+08}, respectively), late-time X-ray and radio-wavelength studies of SNe~Ia/IIn may reveal much about the CSM profile.

\section{CONCLUSION}\label{sec:conclusion}

Spectra of PS15si are well fit by spectra of other SNe~Ia/IIn observed previously \citep{silverman+13}. Comparisons to these examples and some SN~Ia subtypes, however, reveal new underlying diversity in this class of SNe, as follows.

\begin{enumerate}
	\item PS15si is best fit by spectra of overluminous SNe~Ia (e.g., SN~1991T) if we add extra thermal continuum, but there are inconsistencies in the continuum level at different epochs as well as the strength of [Co\III] emission in the final epoch. This spectral morphology matches the interpretation of other SNe~Ia/IIn, although we interpret this similarity as an indication of the added continuum emission and changes in the ionisation state of the visible ejecta layers brought on by CSM interaction. Detailed radiative-transfer models of shocked ejecta illuminated by CSM interaction are needed in order to satisfactorily reproduce SN~Ia/IIn spectra, but PS15si illustrates that the similarity to overluminous SNe~Ia is incidental to spectra of SNe~Ia/IIn and not indicative of the underlying SN.

	\item PS15si appears to have rebrightened over a short timescale at around $85\days$ after discovery. Similar behaviour was also observed by \citet{inserra+16} for SN~1997cy and in other SNe~Ia/IIn where the mechanism was assumed to be clumpiness or steep density gradients in the CSM. Especially at late times, SN~Ia/IIn environments are poorly fit by models assuming uniform CSM. SNe~Ia/IIn may be good candidates for late-time radio and X-ray observations, which can probe the properties of the CSM out to large distances from the progenitor.
\end{enumerate}

\smallskip\smallskip\smallskip\smallskip
\noindent {\bf ACKNOWLEDGMENTS}
\smallskip
\footnotesize

We thank the staffs at the MMT and Keck Observatories for their assistance with data acquisition. Observations using Steward Observatory facilities were obtained as part of the large observing program AZTEC: Arizona Transient Exploration and Characterization. Some observations reported here were obtained at the MMT Observatory, a joint facility of the University of Arizona and the Smithsonian Institution. Some of the data presented herein were obtained at the W. M. Keck Observatory, which is operated as a scientific partnership among the California Institute of Technology, the University of California, and NASA; the observatory was made possible by the generous financial support of the W. M. Keck Foundation. The authors wish to recognise and acknowledge the very significant cultural role and reverence that the summit of Mauna Kea has always had within the indigenous Hawaiian community. We are most fortunate to have the opportunity to conduct observations from this mountain.  

This work is also based in part on observations collected at the European Organisation for Astronomical Research in the Southern Hemisphere, Chile as part of PESSTO (the Public ESO Spectroscopic Survey for Transient Objects) ESO programs 188.D-3003 and 191.D-0935. Operation of the Pan-STARRS1 telescope is supported by NASA under Grant No. NNX12AR65G and Grant No. NNX14AM74G issued through the NEO Observation Program.

C.D.K.'s research is supported by NASA through Contract Number 1255094 issued by JPL/Caltech.  N.S.'s research receives support from NSF grants AST-1312221 and AST-1515559. The supernova research of A.V.F.'s group at U.C. Berkeley is supported by Gary \& Cynthia Bengier, the Richard \& Rhoda Goldman Fund, the Christopher R. Redlich Fund, the TABASGO Foundation, and NSF grant AST-1211916.

\bibliography{ps15si}

\end{document}